 \documentclass[12pt]{article}
\usepackage{epsfig}
\makeatletter
\oddsidemargin   - .1 in
\evensidemargin   - .1 in
\leftmargin - .1 in
\newcommand{\doublespacing}{\let\CS=\@currsize\renewcommand{\baselinestretch}
{1.0}\tiny\CS}

\newcommand {\beq} {\begin{equation}}
\newcommand {\al} {\alpha}
\newcommand {\eeq} {\end{equation}}
\newcommand {\bt} {\beta}
\newcommand {\om} {\omega}
\newcommand {\dl} {\delta}
\newcommand {\Dl} {\Delta}
\date{}

\begin{document}
\thispagestyle{empty} \setcounter{page}{1} \vskip40pt
\centerline{\bf Dynamic Multiple Scattering, Frequency Shift and}
 \centerline{\bf Possible Effects on Quasars Astronomy}
 \vskip20pt
 \centerline{\footnotesize Sisir Roy$^{1,2}$, Malabika Roy$^1$, Joydip Ghosh$^1$ \& Menas
Kafatos$^1$} \vskip20pt \centerline{\footnotesize $^1$Center for
Earth Observing and Space Research, School of Computational
Sciences,} \centerline{\footnotesize  George Mason University,
Fairfax, VA 22030  USA} \centerline{\footnotesize $^2$Physics and
Applied Mathematics Unit, Indian statistical Institute, Kolkata,
INDIA}
 \vskip15pt \noindent
\footnotesize{e.mail: $^1$ mkafatos@crete.gmu.edu} \\
\footnotesize{e.mail: $^1$ sroy@scs.gmu.edu} \\

\vskip15pt
 \abstract{\noindent{\footnotesize{ The shifting of spectral lines due to induced
  correlation effect, discovered first by Wolf for the single scattering case which mimics
  the Doppler mechanism has been extended and developed further by the present authors
   to study the behavior of spectral lines in the case of
multiple scattering and observed shifting, as well as broadening of
the spectrum. We have explored Dynamic Multiple Scattering(DMS)
 theory for explaining anomalous redshifts in quasars. Our
recent work, based on the statistical analysis of the V\'eron-Cetty
data(2003) supports that quasar redshifts fit the overall Hubble
expansion law, as in the case of galaxies, for $z\leq 0.295$ but not
for higher redshifts, indicating  clearly  the inadequacy of the
Doppler effect as the sole mechanism in explaining the redshifts for
high redshift quasars($z\geq 0.295$). We found that the redshift
posseses an additive,``discordant''
 component due to frequency shifting from the correlation induced
 mechanism which increases
gradually for $\sim 0.295 \leq z \leq 3.0$, however, appearing to
follow the evolutionary picture of the universe with absolute
dependence on the physical characteristics i.e., environmental
aspects of the relevant sources through which the light rays pass,
after being multiply scattered. According our framework, as the
environment around sources is diverse, subject to the age of the
universe, it determines the amount of multiple scattering effect,
probably, without additional additive effects for higher values
i.e., for redshifts \ $ 3.00 \leq z$. The recent observational data
on redshift $z$ versus apparent magnitude($m$) (Hubble like
relation) are found to be in good agreement, considering suitable
values of the induced correlation parameters. This resolution of the
paradox of quasar redshifts is much more appealing and in a sense,
more mainstream physics than either assigning redshift entirely to
the Doppler effect or inventing a new, often unknown, physical
mechanism. Our analysis indicates the importance of local
environmental aspects of relevance (recent observations of molecular
gases, the plasma like environment, evolution of the hydrogen
content with epoch etc.) around quasars, especially for the higher
redshift limits. Our work opens possible new vistas in quasar
astronomy as well as for cosmological models of the universe.}}}
\vskip10pt \noindent
 {\bf {keywords}} Dynamic Multiple Scattering, Doppler shift, Quasars, Hubble law
\vskip10pt \noindent
 [Running Title: Dynamic Multiple scattering]
\newpage
\section{{\bf Introduction}}

The frequency shift of spectral lines is most often attributed to
the Doppler effect, and the Doppler broadening of a particular line
 depends on the temperature, pressure or the
different line of sight velocities. In the last few years, a dynamic
multiple scattering theory$^{1,2}$ has been developed to account for
the shift as well as the broadening of spectral lines, even in the
absence of any relative motion between the source and the observer.
This development followed the theoretical prediction made by
Wolf$^{3,4}$ who discovered that under appropriate circumstances, it
is possible to have frequency shifts of spectral lines which
typically mimic all the characteristics of the frequency shifts
obtained by the Doppler mechanism. Since then, it has been verified
by experiments$^{5,6}$ that, contrary to a commonly held belief, the
spectrum of light is, in general, not invariant on propagation
through a medium, especially, for anisotropic (i.e., inhomogeneous)
medium with fluctuations of the dielectric susceptibility, both
temporal and spatial. For example, if radiation is scattered by some
scatterer, the spectrum may change; its maximum may shift either to
the red or to blue part of the spectrum, depending upon the physical
parameters involved in the process. \vskip5pt \noindent It is well
known fact that at higher temperatures, more energy is concentrated
in the high-frequency part of the spectrum(Planck distribution).
According to Wolf$^{3,4,7}$, this distribution is not universal: \
the spectrum can change if the source is partially coherent with
respect to the spatial and temporal coordinates. Wolf$^{3,4}$, and
subsequently  others examined about the possible
implications$^{1,2,8-15}$. These conditions are generally found in
the case of light passing through an anisotropic, weakly
turbulent(tenuous) i.e., some kind of inhomogeneous medium. In such
case, the shift as well as the broadening of the spectral lines can
be caused by the induced correlation mechanism within the
medium$^{16}$. An intuitive understandings of this effect has been
elaborated further by Tatarskii$^{17}$. Following this framework,
sufficient conditions for redshift (i.e., when the shifting is
larger than broadening), also known as no-blue shift
condition$^{18}$, have been derived by Datta et al.$^{1,2}$. They
also calculated the critical condition for the source
frequency$^{19}$ under which the spectrum will be analyzable without
blurring. Recently, it has been pointed out by several
authors$^{11-13,19-21}$ that this mechanism may explain the
anomalies in observed redshift measurements especially in the case
of quasar astronomy. It is generally believed that the medium around
a quasar is a highly anisotropic and plasma like (i.e., can be
considered as underdense). \vskip5pt \noindent The paper is
organized as follows : before entering into the details of the
problem, we describe, as a background, the correlation induced
mechanism and dynamic multiple scattering in section II  and III,
respectively. Section IV deals with the impact of the environment on
the incident radiation, emerging from of a source, on the spectral
line width of the spectrum traversing the medium and consequently on
redshift observations. The width of the spectral line is calculated
after multiple scatterings and the contribution of this mechanism on
the variation of the width with redshift ${z}$ is studied in quasar
astronomy, especially, in the case of quasars where the nature of
observed variation is difficult to explain. Using the relation of
width and ${z}$, \ Roy et al.$^{(8)}$ has already established a
generalized relation  for distance modulus which can be reduced to a
Tully-Fisher type relation, as the contribution due to Wolf effect
becomes almost negligible for low redshifts. Ultimately, this points
to the possible impacts on the Hubble diagram itself, especially,
when we approach higher redshifts ($z \geq 0.3$). \vskip5pt
\noindent The Hubble flow for the standard cosmological models is
considered and a comparison is made with the Hubble Law, following
our framework and also the statistical fitting based on the
nonparametric statistical analysis and regression. The clear
deviation of the linear Hubble law for high redshift ($z$) quasars
is explained in our framework for a certain range of medium
parameters i.e., $\alpha, \alpha^\prime$ and $\beta$. However, this
kind of deviation (or the existence of a {\it bulge}) has already
been mentioned by several authors in connection with the Hubble
diagram for high redshift Supernovae$^{(22),(23)}$. Our analysis
clearly indicates that, in addition to observed redshifts, caused by
the Doppler
 and/or gravitational effects, some  quasars may contain contributions due
 to presence of the induced correlation properties inside the fluctuating medium,
arising from intrinsic properties of the medium surrounding the
radiating sources. Finally, in section V and VI, we discuss possible
implications and conclusions regarding  the application of the
presently discussed mechanism to quasar astronomy and cosmology in
general. \noindent \vskip10pt
\section{Correlation Induced Mechanism : Nature of the Source and the Medium }

\subsection {Dielectric Response in a Random Medium with spatial and Temporal Fluctuation}
The scattering of electromagnetic waves by a random medium involves
the solution of Maxwell's equations and an ensemble average over
random fluctuations. Such calculations can be carried out for the
first order Born approximation. Recently a Japanese group$^{14,15}$
showed that the higher order Born approximations contribute of
negligible amount in case of multiple scattering in a random,
tenuous medium. In this case, i.e., in the first order Born
approximation, the medium is treated as tenuous which means the
filling factor, i.e., the ratio of the volume occupied by particles
to the total volume of the medium, is considerably smaller than $0.1
\%$. When it becomes much greater than $1\%$, the problem is to be
treated as diffusion, but in the neighborhood of $1\%$ neither of
the two treatments is valid and it requires a solution considering
the complete equation of transfer. Thus the present case is for a
medium in which the dielectric constant, permeability and the
conductivity are all random functions of position and time. Strict
homogeneity requires, of course, that the medium occupies all space.
However, in the present case, we assume that the linear dimensions
of the domain are very large compared to the distances over which
the correlations of the fluctuations in the macroscopic physical
properties of the medium are non-negligible. Also, we consider the
dielectric response in a very weakly fluctuating random medium.
\vskip5pt \noindent Let us begin by recalling some standard
relations between the electric field and the induced polarization in
a linear inhomogeneous, isotropic and nonmagnetic medium. At the
end, we will deal with the associated magnetic field too. The
Fourier transforms are given by
$$\tilde{\bf E}({\bf r} \omega) = \frac{1}{2\pi}\int^\infty_\infty {\bf E}({\bf r},t)e^{(i\omega t )}dt$$
$$\tilde{\bf P}({\bf r},\omega) = \frac{1}{2 \pi}\int^\infty_\infty {\bf P}({\bf r},t)e^{(i\omega t )}dt$$
of the real electric field ${\bf E}({\bf r},t)$  and the induced
polarization ${\bf P}({\bf r},t)$ respectively$^{24}$. Here, ${\bf
r}$ denotes the position vector and $t$ the time. ${\bf E}({\bf
r},t)$ and ${\bf P}({\bf r},t)$ are related by a constitutive
relation
\begin{equation}
\tilde{\bf P}({\bf r},\omega) =  \tilde\eta ({\bf
r},\omega)\tilde{\bf E}({\bf r} \omega)
\end{equation}
 where $\tilde \eta ({\bf r}$ denotes as the dielectric
 susceptibility, describing the
 response of the medium at the frequency $\omega$ of the incident wave. According to
microscopic theory,  $\tilde\eta ({\bf r},\omega)$ can be expressed
in terms of the average number ${\cal N}({\bf r})$ of molecules per
unit volume present in the medium and the mean polarizability
$\alpha(\omega)$ of each molecule by the Lorentz-Lorenz relation,

\begin{equation}
\tilde\eta ({\bf r},\omega) = {\bf {\cal N}}({\bf
r})\alpha(\omega)[1-(\frac{4 \pi}{3}){\bf {\cal N}}({\bf r})
\alpha(\omega)]^{-1}
\end{equation}
Let the dielectric constant $\epsilon_r$ be described by a random
function of position and time
\begin{equation}
\epsilon_r  = \epsilon({\bf r}, t) = n^2({\bf r}, t)
\end{equation}

which in turn, is related with the dielectric susceptibility $\tilde
\eta ({\bf r}$ as
\begin{equation}
n^2(\omega) = 1 + 4 \pi \tilde{\eta}(\omega) \end{equation}
 implying
\begin{equation}
\frac{4 \pi}{3} {\cal N} \alpha(\omega) = \frac{n^2(\omega -
1)}{n^2(\omega +2)}\end{equation} \vskip5pt \noindent The
polarizability \ $\alpha(\omega)$ \ is a causal, complex function of
frequency and, consequently, its real and imaginary parts are
coupled by dispersion relations$^{25}$. It contains the frequency
dependence and plays a crucial role, especially, when dealing with
the scattering of light waves in a plasma medium alone. In the
present case, ${\cal N({\bf r}},t)$ is assumed to be real,
stationary, homogeneous random function of position and time, of
zero mean value. For dilute gas, ${\cal N({\bf r}},t)$ \ represents
the density of molecules at the point ${\bf r}$ \ and time $t$,
while $\alpha(\omega)$ \ represents the response of the individual
molecules to the applied field oscillating with frequency $\omega$.
The scattering from the plasma medium has been extensively studied
by various groups using the first Born approximations.
Watson$^{26,27}$, studied the multiple scattering of electromagnetic
waves both in dense and tenuous plasmas. However, it is not trivial
to relate the dielectric constant to the plasma properties. In the
case of plasma, coherently interfering wavelets propagating in a
medium with a refractive index $n$, can be expressed in terms of
plasma correlation functions and density. On the other hand, the
coherent scattering is usually described by "classical like"
transport equations$^{28,29}$. For the present case, we are dealing
with the physical characteristics, as described here, without going
into the full details of the plasma case. \vskip5pt \noindent
 It is to
be noted here that as there is substantial velocity of gas molecules
or other particles present in the medium, and the velocity is a
vector field , we need to construct a general structure function
between different components. Also, we need to consider the effects
of pressure, partial pressure, temperature etc. of a particular
component like hydrogen. However, we take the Fourier transform of
Eq.(1)  and make use of the fact that the response of the medium is
necessarily causal. Then, we can write the following constitutive
relation :
\begin{equation}
{\bf P}(t) = \frac{1}{2 \pi}\int^t_{-\infty} \eta(t - t^\prime){\bf
E}(t^\prime)dt^\prime
\end{equation}
or, equivalently
\begin{equation}
{\bf P}(t) = \frac{1}{2 \pi}\int^\infty_{0} \eta(t^\prime){\bf E}(t
- t^\prime)dt^\prime \end{equation}
 where,

\begin{equation}
\eta(\tau) = \int^\infty_{-\infty} \tilde{\eta}(\omega) e^{-i \omega
\tau} d\omega
\end{equation}
\noindent In our case,  we consider a medium where, \ $\eta$ \
depends on time through \ ($t- t^\prime$) instead of $t$ or
$t^\prime$ and also $\eta(\tau) = 0$ for $\tau <
0$ because of the assumption of causality. \\
If the response of the medium is time dependent, the number density
of the medium also changes with time and consequently, if temporal
variation are not too strong , we have
\begin{equation}
\hat \eta({\bf r},t;\omega) = \frac{{\cal N}({\bf
r},t)\alpha(\omega^\prime)}{1 - \frac{4 \pi}{3} {\cal N} ({\bf
r},t)\alpha(\omega^\prime)}
\end{equation}
Though the precise range of validity of the above relation is
heavily dependent on the microscopic considerations, it clearly
points to the physical origin of the generalized dielectric
susceptibility which depends on two rather than one temporal
arguments, i.e., $\eta({\bf r},t;\omega)$ \ is a random function
with respect to its first time argument and deterministic function
with respect to its second time argument. \vskip5pt \noindent
 For practical purposes, when the effective frequencies(say $\omega_0$)
\ of the electric field are not too close to any of the resonance
frequencies of the medium, the random fluctuations of the scattering
medium are taken to be stationary, at least in the wide sense$^{30}$
and the medium is also homogeneous so that we can arrive at an
approximated formula as follows$^{31}$:
\begin{equation}
{\bf P} ({\bf r}, t) = \hat \eta({\bf r},t;\omega_0){\bf E}({\bf
r},t)
\end{equation}
It is this approximate nature of time dependent response
 function $\hat\eta(\bf r,t)$ which is responsible for
 determining its validity in  the field which contains frequencies,
 near or far away  from resonances. So, in this correlation
 induced mechanism, we have considered the random susceptibility $\eta$ as a function of
$\omega,{\bf r},t$, and under these conditions, $\eta$ \ is of the
approximate form
\begin{equation}
\eta(\omega,{\bf r},t) = \eta_1(\omega) \eta_2({\bf r},t)
\end{equation}
Under suitable circumstances, one can be consider \ $\eta(\omega)
={\rm constant}$, \ as effectively independent of $\omega$. This is
analogous to the dynamic structure factor for particle density
fluctuations$^{28}$.
\subsection{Condition for the Change of initially incoherent radiation to a
coherent spectrum after scattering } \vskip5pt\noindent In
discussing the propagation and scattering of an initially incoherent
radiation, let us first try to understand what circumstances may
change the initial spectrum. Let us consider, at first, that an
initial spectrum is from an incoherent source. Now, some scatterer
may be present having its size small compared to that of the source,
at an arbitrarily large distance. Consequently, the scatterers can
be considered as illuminated practically by a spatially coherent
radiation. Electromagnetically, it can be described as a scattering
phenomenon where radiation from the secondary dipoles fall on the
scatterer under the influence of the incident wave. As the coherent
source may distribute its energy in different directions,
differently at different frequencies, then, in some directions, the
scattered radiation may have a quite different (from the initial)
distribution of the energy in frequency space, i.e., the Wolf shift
may appear, as has also been concluded by Tatarsky$^{17}$. Then, if
for some reason, we do not observe the direct radiation(say, the
angular distance between the primary source and scatterer is very
large), but only a scattered one, we may conclude that the spectrum
corresponds either to a red shift or a blue shift. In other words,
an initially, spatially incoherent field becomes partially coherent
at some distance, and the coherence radius increases with distance.
This phenomenon is closely related to the theorem of Van
Cittert$^{32}$ and Zernicke$^{33}$. \vskip5pt \noindent
 Also, in all wave fluctuation phenomena, irrespective of
the form or nature of correlation function, if the correlation
distance is much smaller than the Fresnel zone size $\sqrt{\lambda
L}$, where $\lambda= {\rm wavelength}$ \ and \ $L$ is distance of
the far zone of scattering, then in general, the amplitude and phase
variances are equal and proportional to the square of frequency and
to the distance $L$. But, if the correlation distance is much larger
than the Fresnel zone, then the cross section becomes independent of
frequency and proportional to $L^3$. Physically it means that for a
short distance, the amplitude does not vary much but the phase does
with the distance of wave propagation. But as the distance
increases, the amplitude fluctuation also increases and practically
there does not remain any difference between amplitude and phase
fluctuations. \vskip10pt
\subsection{Effects of Fluctuation on the Behavior of the Field in the Medium}
\vskip5pt \noindent  Many physical characteristics of the medium
determine the behavior of the incoming or interacting field. One
among them is the optical distance \ $\gamma = \rho \ \sigma_t \ x$,
\ where, \ $\rho$ \ is the density of particles, \ $\sigma_t$ \
denotes the total cross section and $x$ the semi-infinite region \
$(x>0)$$^{34}$, \ respectively. For $\gamma << 1$, \ the field
becomes predominantly coherent whereas, when $\gamma \simeq 1$, \
the opposite behavior arises (incoherent or fluctuating field). This
causes an amplitude fluctuation behavior. However, we consider a
medium with weak fluctuation where the field is predominantly
coherent i.e., the degree of coherency is much more than that of
incoherency, which means \ $\gamma
>> 1$. \ Another case can also be encountered in practice when
the angle of scattering is very small and the role of coherent
field(average field) dominates even with optical distances
comparable to unity. In such a situation, the field can be
considered as weakly fluctuating as the fluctuation can be taken as
a stationary process which means the mean period and correlation
time to be much shorter than the averaging interval needed to make
the observation. \vskip5pt \noindent
 Now, when an electromagnetic wave is incident
upon a turbulent medium, the amplitude as well as the phase of the
incident wave experience fluctuations due to the  relative
dielectric constant $\epsilon_r$ or the index of refraction \ $n$
which can vary both spatially and temporarily in the case of
turbulent medium (through Eq.(3) and (7)). According to the present
framework, the medium around the radiating source is generally
characterized by the permittivity $\epsilon({\bf r},t)$, which is a
random function not only of space and time but also of other
characteristics like pressure, temperature density of particles,
degree of ionization etc., through the relation with refractive
index, as explained earlier. \vskip5pt \noindent
 It is evident from Eq.(2,3 \& 4 ) that in any linear
medium, the dielectric susceptibility can also be related to the
dielectric co-efficient. Now, in the case of a random but weakly
fluctuating medium, we can write
\begin{equation}
\epsilon({\bf r},t) = <\epsilon({\bf r},t)> (1 + \epsilon_1({\bf r},
t))
\end{equation}
\noindent where, $\epsilon_1({\bf r}, t)$ \ is the fluctuating part
of the permittivity with  \ $<\epsilon_1> = 0$. Also, the refractive
index $n({\bf r}, t)$ \ being a random function of space and time
can be written as
\begin{equation}
n({\bf r}, t) = \sqrt {\frac{\epsilon({\bf r}, t)}{\epsilon_0}} =
<n({\bf r}, t)> [1 +n_1({\bf r},t) ]
\end{equation}
\noindent $\epsilon_0$ \ being the free space permittivity, assumed
to be constant and $n_1$, the
 fluctuating part of the refractive index. For small
 fluctuations i.e., considering a weakly fluctuating medium, we obtain
\begin{equation}
 \epsilon_1({\bf r},t) \simeq 2 n_1({\bf r},t)
\end{equation}
\noindent  When considering an ionized plasma medium, such as the
ionosphere, it can be assumed $^{34}$
\begin{equation}
\epsilon_1 \simeq g\frac{\delta \rho}{\rho}
\end{equation}
where,
$$g = -\frac{\omega_{N_{\rm e}}^2}{\omega^2};  \ \ \ \
\omega_{N_{\rm e}}^2 = (\frac{1}{\epsilon_0}) (\frac{N_{\rm e}
e^2}{m_e})$$ \noindent $N_{\rm e}$ is the number density of  free
electrons in the medium and $m_e$ \ being the mass of the electron.
$\epsilon_1$ \ is assumed to be due to the fluctuations in electron
concentration, proportional to the density $\Delta\rho$ of the
carrier medium.  Villars and Weisskopf$^{34}$ studied the scattering
of electromagnetic waves in the ionosphere, with such an
$\epsilon_1$ and showed that the scattering is compatible with
observations. Therefore, for the case of small fluctuations,
\begin{equation}
2 n_1 \simeq g \frac{\delta \rho}{\rho}
\end{equation}
 Taking the average,
\begin{equation}
<n_1^2> \simeq \frac{g^2}{4}<\frac{\delta \rho}{\rho}\frac{\delta
\rho}{\rho}> = \phi(\omega). f({\bf r},t)
\end{equation}
\noindent Here, the time-dependence in $\delta \rho$ is assumed to
lie outside the domain of propagated frequencies $\omega$,
considering \ $\phi$ \ is a function of frequency $\omega$ and,
 taking $f$ \ as a function of space and time only. In this way, even if, one considers
 the scattering from ionized plasma but, far away from resonance region, the correlation
 induced mechanism can be applied$^{24,35}$. Likewise, we assume that the
correlation of refractive indices at two points in the ionized
plasma medium (tenuous and underdense) can also be decomposed into
two functions , a function of frequency and a function of space and
time. It is to be mentioned, for the real value of  $n^2$, there
will be a sharp cut-off at the resonance. Otherwise, they will be
diffused. There also will be a cut-off when the phase velocity
approaches infinity and correspondingly $n^2 = 0$ \ e.g., at low
frequencies , \ $\omega = \omega_{re}$. \ In case of resonance, the
phase velocity will be zero and $n^2 \rightarrow \infty$. \ Normally
the sharpness condition is applied to collisionless plasma, but
collisions can also be taken into account under certain
conditions$^{35}$.
 \vskip5pt
 \noindent
 However, we also need to consider other factors, such as, the dependence
 on the distance between source and scattering medium, the spatio-temporal
 fluctuation of structure functions, together with other characteristics of the
medium, like, densities of different particles present, magnetic
field strength leading to the variation of degree of ionization,
etc. Now, within certain distance limit, when the intensity of the
turbulence is weaker than a certain level, some mathematical
simplifications are possible and the case can be referred as "weak
fluctuation" approximation. \vskip5pt \noindent In this way, many
processes encountered in practice can be approximated reasonably
well within the domain of stationary or homogeneous random
functions, even if, the approximation is only valid within a limited
time or spatial distance. Mathematically, this kind of physical
processes when treated as random functions are referring to a random
process with stationary increments of a function of time and locally
homogeneous random function of position and the description is done
in a convenient way by taking a two point correlation function
analogous to the so called Van Hove time dependent two-point
correlation function$^{28,30,31}$ \ in neutron scattering, the
medium being treated statistically in the behavior and consequences
of the wave scattering.
\section{\bf Dynamic Multiple Scattering}
The scattering of light induced by the random susceptibility which
can produce both a shift and a broadening of the spectral lines,
other than Doppler mechanism, has been studied in statistical
optics$^{35}$, where the existence of a random dielectric
susceptibility fluctuation, both spatially and temporally, is
assumed. This is the so called  Wolf effect$^{3,4}$. Following the
above formalism, during the last decade, some closely related
processes have been developed that can generate frequency shifts of
spectral lines$^{1,2,9-13,18}$. Such spectral effects were first
investigated in $^{3,4}$ for the case of a static scattering medium;
the analysis for the case of time-dependent scatterers followed soon
thereafter$^{14,15}$. To start with, we briefly present the main
results of Wolf's scattering mechanism. \vskip5pt \noindent

\subsection{Single Scattering by Wolf Mechanism : Wolf Effect}
\vskip5pt \noindent
 Let us consider the case when both the source and the medium
 through
 which scattering occurs, to be randomly fluctuating. Together with
 this, we assume a polychromatic electromagnetic field of light of central
 frequency $\omega_0$ and width $\delta_0$, incident on the scatterer.
 The general dielectric susceptibility $\eta$, considered here, characterizes the response of the
  space-time fluctuating medium, and is given as a function of three variables : \ the position
   vector ${\bf R}$ and the time \ $\tau$ resulting from, i.e., thermal fluctuations of the medium and the
   frequency \ $\omega$ \ of the incident field. More specifically, the correlation function
   is taken as homogeneous and stationary, at least in a general sense and with a zero mean,
   a generalized analogue of the Van Hove two-particle correlation function$^{28}$, frequently
   encountered in neutron scattering. Let the incident spectrum be assumed
   to be of the form$^{36}$
\begin{equation}\displaystyle{
S^{(i)}(\omega)=
A_0e^{\left[-\frac{1}{2\delta_0^2}(\omega-\omega_0)^2\right]} }
\end{equation}
The spectrum of the scattered field is given by$^{(12)}$
\begin{equation}\displaystyle{
S^{(\infty)}r({\bf u'},\omega')=A\omega'^4\int_{-\infty}^{\infty}
{\cal K}(\omega,\omega',{\bf u}, {\bf u'})S^{(i)}(\omega)d\omega}
\end{equation}
which is valid within the first order Born approximation$^{36-38}$.
$A$, the spectral amplitude after being scattered, can be determined
in terms of $A_0$, by
$$A_0 = \frac{(2\pi)^3 V (\sin^2 \psi)}{c^4 r^2}$$
where, $V$ \ represents the volume of the scattering medium (whose
linear dimension is assumed to be large compared with the
correlation width of the medium) \ and \ $\psi$  denotes the angle
between the direction of the electric vector of the incident wave
and the direction of the scattering, $c$ is the speed of the light
in vacuum.  ${\bf u}$ and ${\bf u'}$ are the unit vectors in the
direction of incident and scattered fields respectively. Here,
${\cal K}(\omega^\prime, \omega;\textbf{u}^\prime,\textbf{u})$ \ is
the so called scattering kernel and plays the most important role in
this mechanism, given by
\begin{equation}
{\cal K}(\omega^\prime, \omega; {\bf{u}^\prime} , {\bf{u})} \equiv
 \widetilde{G} \left (\frac{\omega^\prime}{c}{\bf u}^\prime - \frac{\bf{u}}{c} \bf{u},
  (\omega^\prime - \omega);\omega \right  )
\end{equation}
\noindent with
$$\widetilde{G}(\textbf{k},\Omega; \omega) =
\frac{1}{(2\pi)^4} \int_V d^3 R \int^\infty_\infty dT
G(\textbf{R},T;\omega) \times exp[-i(\textbf{k}. \textbf{R} - \Omega
T)]$$

\vskip5pt \noindent As the dielectric susceptibility is considered
in a statistically homogeneous and stationary medium, the angular
brackets in the correlation function \ $<\hat \eta^*({\bf
r_1},t,\omega^\prime)\hat \eta({\bf r_2},t^\prime,\omega^\prime)>$ \
denote the average over an ensemble of the random medium depending
on ${\bf {r_1}}$, ${\bf {r_2}}$ and $t, \ \ t^\prime$ only through
the difference ${\bf {(r_1 - \bf r_2)}}$, and $(t^\prime-t)$, \
respectively. Also, the medium where the character of the
fluctuations does not change with time, even though any realization
of the ensemble \ $x(t)$ \ changes continually in time, is said to
be statistically stationary. It means that all the ensemble averages
are independent of the origin of time; \ moreover, the field is as a
rule also "ergodic", i.e., may be called "macroscopically steady".
Therefore, the scattering medium introduced here, possesses
space-time fluctuations in which case we assume the random
fluctuations of the generalized dielectric susceptibility as
"ergodic" i.e., it obeys Gaussian statistics with a zero mean. Then,
instead of studying $\cal K(\omega,\omega')$ \ in detail, we
consider a particular case for the correlation function
$G({\bf{R}},T;\omega)$ of the generalized dielectric susceptibility
$\eta({\bf r},t;\omega)$ of the medium, characterized by an
anisotropic Gaussian function
\begin{equation}\displaystyle{
\begin{array}{lcl}
G({\bf{R}},T;\omega^\prime) & = &
<\eta^*({\bf{r}}+{\bf{R}},t+T;\omega)\eta({\bf{r}},t;\omega)>\\ \\
& = &G_0exp\left[-\frac{1}{2}\left(\frac{X^2}{\sigma_x^2}
+\frac{Y^2}{\sigma_y^2}
+\frac{Z^2}{\sigma_z^2}+\frac{c^2T^2}{\sigma_\tau^2}\right)\right]
\end{array}
}
\end{equation}

\vskip5pt \noindent where, $~G_0~$ is a positive constant,
$~{\bf{R}}~=~({\bf{r_2}}-{\bf{r_1}}) = ({\bf X,~Y,~Z~,}) \ {\rm and}
 \ \tau = (t^\prime - t)$. \
$~\sigma_x~,~\sigma_y~,~\sigma_z~,~\sigma_\tau~$ are correlation
lengths representing effective distances over which the fluctuations
of the dielectric constant are correlated. The anisotropy is
indicated by the unequal correlation lengths in different spatial as
well as temporal directions. However, we assume the \ $\sigma$' s \
to be much smaller than the linear dimensions of the scattering
volume. With these considerations, \ $\cal K(\omega,\omega')$ \ can
be obtained from the four dimensional Fourier Transform of the
correlation function $G({\bf{R}},T;\omega^\prime) $. \vskip5pt
\noindent Here, the spatial extent of the susceptibility correlation
is sufficiently small compared to the size of the scattering volume,
$V$. This means the separation distance $R=|{\bf R}|=|{\bf {r_2}-\bf
{r_1}}|$ for which $|G({\bf R},T; \omega^\prime)|$ \ assumes
appreciable values, is much smaller than the linear dimension of
$V$. In general, the three correlation lengths $\sigma_x$, \
$\sigma_y$ and $\sigma_z$ \ differ from each other and the medium is
then statistically anisotropic. The correlation function (eq. 21) is
a natural generalization of those considered in refs.$^{38,39}$.
Thus when the Born approximation is applied, the cross section can
always be well estimated in terms of suitably chosen generalized
time dependent pair distribution function $G({\bf {R}}, T)$ from the
statistical point of view. This can also be stated the average
density distribution at a time \ $t^\prime + t$, \  seen from a
point where the particle passed at time $t^\prime$ $^{27}$. The
Fourier transform, defined by equation(20), of $G(\textbf{R}
T;\omega)$  is readily found to be

$$\widetilde{G}(\textbf{k},\Omega; \omega) = B exp[\sigma_x^2 k^2_x +
\sigma_y^2 k^2_y + \sigma_z^2 k^2_z + \sigma_\tau^2\Omega^2/c^2]$$
\noindent in which
$$B = \frac{G_0 \sigma_x \sigma_y \sigma_z \sigma_\tau}{c(2 \pi)^2}$$
\noindent and $(k_x,k_y k_z)$ \ are the Cartesian components of the
vector $\textbf{k}$ \ with respect to the same coordinate axes as
chosen for $\textbf{R}$. \vskip5pt \noindent
 Or, \ ${\cal K}(\omega,\omega';{\bf u},{\bf u^\prime})$ \ can be shown to be of the form
\begin{equation}\displaystyle{
{\cal{K}} (\omega,\omega')=B \ exp\left[-{\frac{1}{2}}
\left(\alpha'\omega'^2- 2\beta\omega\omega'+\alpha\omega^2 \right)
\right] }
\end{equation}
where
\begin{equation}\displaystyle{
\left.
\begin{array}{lcl}
\alpha &=&
{\frac{\sigma_x^2}{c^2}}u_x^2+{\frac{\sigma_y^2}{c^2}}u_y^2+
{\frac{\sigma_z^2}{c^2}}u_z^2+{\frac{\sigma_\tau^2}{c^2}}\\ \\
\alpha' &=&
{\frac{\sigma_x^2}{c^2}}u_x'^2+{\frac{\sigma_y^2}{c^2}}u_y'^2+
{\frac{\sigma_z^2}{c^2}}u_z'^2+{\frac{\sigma_\tau^2}{c^2}}\\ \\
{\rm and} \ \  \ \ \ \ \beta &=&
{\frac{\sigma_x^2}{c^2}}u_xu_x'+{\frac{\sigma_y^2}{c^2}}u_yu_y'+
{\frac{\sigma_z^2}{c^2}}u_zu_z'+{\frac{\sigma_\tau^2}{c^2}}
\end{array}
\right\} }
\end{equation}
Here $\hat{u}=(u_x,u_y,u_z)$ and $ \hat{u'}=(u_x',u_y',u_z')$ are
the unit vectors in the directions of the incident and scattered
fields respectively.
 \vskip5pt \noindent Substituting (22) \& (23)
in (19), we obtain
\begin{equation}\displaystyle{ S^{(\infty)}(\omega')=
A'e^{\left[-\frac{1}{2\delta_0'^2}(\omega'-\bar{\omega}_0
)^2\right]} }
\end{equation}
where
\begin{equation}\displaystyle{
\left.
\begin{array}{lcl}
\bar{\omega}_0 &=& \frac{|\beta|\omega_0}{\alpha'+\delta_0^2
(\alpha\alpha'-\beta^2)}\\ \\
\delta_0'^2 &=& \frac{\alpha\delta_0^2+1}{\alpha'+\delta_0^2
(\alpha\alpha'-\beta^2)}\\ \\
 {\rm and} \ \ A' &=& \sqrt{\frac{\pi}{2(\alpha\delta_0^2+1)}}ABA_0{\omega^\prime}^4\delta_0
 exp\left[\frac{|\beta|\omega_0\bar{\omega}_0-\alpha\omega_0^2}
{2(\alpha\delta_0^2+1)}\right]
\end{array}
\right\} }
\end{equation}

Though $A'$ depends on $\omega'$, it was approximated by James and
Wolf$^{(9),(10)}$ to be a constant so that $S^{(\infty)}(\omega')$
can be considered to be Gaussian.
 \vskip5pt \indent
 The multiplicative factor \ ${\omega^\prime}^4$, \ just
as in the case of Rayleigh scattering, produces a small amount of
blue shift which is frequency-dependent. It also produces a
frequency dependent change of line intensity  which results in the
source appearing more bluer or otherwise. In other words, it causes
a slight distortion of the Gaussian line spectrum. However, it is
small enough to be neglected for practical purpose. Let us now
define the relative frequency shift as
\begin{equation}\displaystyle{
z=\frac{\omega_0-\bar{\omega}_0}{ \bar{\omega}_0} }
\end{equation}
where $\omega_0$ and $\bar{\omega}_0$ denote the  unshifted and
shifted central frequencies respectively. The spectrum is
redshifted[$~z~>~0$~] or blueshifted [$z~<~0~$] according to the
expression, in terms of relative contribution from the correlation
parameters, given by

\begin{equation}
\displaystyle{
z=\frac{\alpha'+\delta_0^2(\alpha\alpha'-\beta^2)}{|\beta|}-1 }
\end{equation}
It is important to note that this $z$-number does not depend on the
incident frequency, $\omega_0$. As a result, such changes in the
spectrum are similar to the Doppler shifts when the $\omega^4$ \
factor is reasonably omitted. The phenomenon of the changes in this
spectrum does not pertain to the classes of Raman, Brillouin and
other nonlinear scattering. It is a very important aspect which can
have viable applications in the astronomical domain. Expression (27)
implies that the spectrum can be shifted to the blue or to the red,
according to the sign of the term
$\alpha'+\delta_0^2(\alpha\alpha'-\beta^2)~>~ |\beta|~$. \vskip5pt
\noindent
 To obtain the no-blueshift condition $^{(18)}$, we use
Schwarz's Inequality which implies that
$$\alpha\alpha'-\beta^2~\geq~0~$$
Thus, we can take
$$\displaystyle{
 \alpha'~>~|\beta|
 }$$
as the sufficient condition to have only redshift in this mechanism.
\vskip5pt

\subsection{\bf Multiple Scattering}
\vskip5pt\noindent Let's now assume that the light in its journey
from its origin of the source to the observer, for example, from an
astronomical object of a high redshift, encounters many such
scatterers. What we observe at the end is the light scattered many
times, with an effect as that discussed above for each individual
process. Here, the random medium can be modeled in the following
manner: \\
Let the medium consisting of thin multiple layers separated by free
space and each layer is supposed to be very weak so as to produce
only a single scattering. The light is incident on the first random
medium, gets scattered from the first medium and then again incident
on the second random medium that is located in the far zone of the
first random medium etc. By repeating this process, multiple
scattering takes place. \vskip5pt \noindent Let there be $N$
scatterers between the source and the observer and $z_n$ denote the
relative frequency shift after the $n^{th}$ scattering of the
incident light from the $(n-1)^{th}$ scatterer, with $\omega_n$ and
$\omega_{n-1}$ being the respective central frequencies of the
incident spectra at $n^{th}$ and $(n-1)^{th}$ scatterers. Then by
definition,
$$\displaystyle{
 z_n =\frac{ \omega_{n-1}-\omega_n}{\omega_n}, ~~~~~~n~=~1,~2,~.~.~.~.,~  N} $$
or,
$$\displaystyle{
 \frac{\omega_{n-1}}{\omega_n}=1+z_n,~~~~~~n~=~1,~2,~.~.~.~.,~ N} $$
Taking the product over $ n$ \ from \ $n ~=~ 1$ \ to \ $n~ =~ N$, \
we get,
$$ \displaystyle{
 \frac{\omega_0}{\omega_ N}=(1+z_1)(1+z_2)~.~.~.~.~.~(1+z_ N)}$$
 The left hand side of the above equation is nothing but the ratio of the
source frequency and the final or  observed frequency $z_f$. Hence,
\begin{equation}\displaystyle{
1 + z_f = (1+z_1)(1+z_2)~.~.~.~.~.~(1+z_N)}
\end{equation}
\vskip5pt \noindent Since the $z$-number due to such effect does not
depend upon the central frequency of the incident spectrum, each
$z_j$ depends on $\delta_{j-1}$ only, not $\omega_{j-1}$ [here
$\omega_j$ and $\delta_j$ denote the central frequency and the width
of the incident spectrum at $(j+1)^{th }$ scatterer]. To find the
exact dependence we first calculate the broadening of the spectrum
after $N$ number of scattering. \vskip5pt
\subsection{\bf Broadening
of spectrum} \vskip5pt \noindent
Using the second equation in (25)
for multiple scattering,
 we can easily write,
\begin{equation}\displaystyle{
\left.
\begin{array}{lcl}
\delta_{n+1}^2 & = &
\frac{\alpha\delta_n^2+1}{\alpha'+(\alpha\alpha'-\beta^2)\delta_n^2} \\ \\
& = &
\left(\frac{\alpha\delta_n^2+1}{\alpha'}\right)\left[1+\delta_n^2\left(
\frac{\alpha\alpha'-\beta^2}{\alpha'}\right)\right]^{-1}
\end{array}
\right\}}
\end{equation}

and, using the first Eq.(25), we can write
\begin{equation}
\displaystyle{ \omega_{n+1} =
\frac{\omega_n|\beta|}{\alpha'+(\alpha\alpha'-\beta^2)\delta_n^2} }
\end{equation}
and then,

\begin{equation}
\displaystyle{ \left.
\begin{array}{rcl}
z_{n+1}&=&\frac{\omega_n-\omega_{n+1}}{\omega_{n+1}}\\ \\
&=&\frac{\alpha'}{|\beta|}\left\{1+\left(\frac{\alpha\alpha'-\beta^2}{\alpha'}
\right)\delta_n^2\right\}-1
\end{array}
\right\}}
\end{equation}
\noindent Now let's assume that the contribution($\chi$) \ due to
DMS to the $(n+1)$ th scattering process to the redshift is very
small, {\it i.e.,}
$$0~<~\chi ~ = ~ z_{n+1}~~<<~1 \ \ \  {\rm for} \ \ \ {\rm all} \ \ \  n$$
Then,
$$\displaystyle{
\begin{array}{lrcl}
& 1+\chi & = &
\frac{\alpha'}{|\beta|}\left\{1+\left(\frac{\alpha\alpha'
-\beta^2}{\alpha'}\right)\delta_n^2\right\}\\ \\
\end{array}}$$
\vskip5pt \noindent In order to satisfy this condition and in order
to have a redshift only ( or positive $z$), we observe that
\begin{equation}
 \left(\frac{\alpha\alpha'-\beta^2}{\alpha'}\right)\delta_n^2~~<<~~1
\end{equation}

Under this condition and from Eq.(29), after neglecting higher order
terms, the expression for $~~\delta_{n+1}^2~~$ can be well
approximated as:
$$\displaystyle{
\delta_{n+1}^2  \approx
 \left(\frac{\alpha\delta_n^2+1}{\alpha'}\right)\left[1-\delta_n^2\left(
\frac{\alpha\alpha'-\beta^2}{\alpha'}\right)\right] }$$ which, after
 simplifying, gives a very important recurrence relation:

$$\displaystyle{
\begin{array}{lcl}
\delta_{n+1}^2 & = & \frac{1}{\alpha'}+\frac{\beta^2}{\alpha'^2}\delta_n^2 \\
\\
& .~. & .~.~.~.~.~.~.~.~.~.~.~.~.~.~  \\ \\
& = & \left(\frac{\beta^2}{\alpha'^2}\right)^{n+1}\delta_0^2
+\frac{1}{\alpha'}\left(1+\frac{\beta^2}{\alpha'^2}+~.~.~.~.~
\frac{\beta^{2n}}{\alpha'^{2n}}\right).
\end{array}
}$$ \vskip5pt \noindent which, for $n = 1,.....N$, becomes
\begin{equation}\displaystyle{
\delta_{ N+1}^2  = \left(\frac{\beta^2}{\alpha'^2}\right)^{{\cal
N}+1}\delta_0^2
+\frac{1}{\alpha'}\left(1+\frac{\beta^2}{\alpha'^2}+~.~.~.~.~
\frac{\beta^{2 N}}{\alpha'^{2 N}}\right). }
\end{equation}
As the number of scattering increases, the width  of the spectrum
obviously changes but the most important topic to be considered is
whether this change of width is below some tolerance limit or not.
This is very important from an observational point of view.
\vskip5pt \noindent Among several techniques of measuring this
tolerance limit, one is the {\it Sharpness Ratio}, defined as
$$ Q=\frac{\omega_f}{\delta_f}$$
where $~\omega_f ~$ \ and \ $~ \delta_f ~$ are the mean frequency
and the width of the observed spectrum. After $N$ number of
scattering, this sharpness ratio, say $Q_N$, is given by the
following recurrence relation :
$$Q_{N+1}=Q_N \sqrt{\frac{\alpha'}{\alpha'+(\alpha\alpha'-\beta^2)\delta_ N^2}-
\frac{1}{\alpha\delta_ N^2+1}}$$ \vskip5pt \noindent where,  the
expression under the square root lies between $0$ \ and \ $1$.
Therefore, $Q_{ N+1}~<~Q_N$, and the line is broadened as the
scattering process goes on. Under the sufficient condition of
redshift [i.e., $~\alpha^\prime >|\beta|~$ ]$^{(2)}$ [details in
ref.2]
\begin{equation}
\Dl\om_{n+1} \gg \dl_n
\end{equation}
\noindent if the following condition holds:
\begin{equation}
\frac{\dl_n\om_0(\al\al'-\bt^2)}{\al'+(\al\al'-\bt^2)\dl_n^2} \gg 1
\end{equation}
$\om_0$ being the source frequency. \vskip5pt \noindent

Relation (34) applies when the  shifting will be more prominent than
the effective broadening so that the spectral lines are observable
as well as analyzable. But, if the broadening is higher than the
shift of the spectral line, the final spectrum will definitely be
blurred and shifting is impossible to detect. So, the relation (35)
is one of the conditions necessary for the observed spectrum to be
analyzable. Again, for large $N$ (i.e., $ N \rightarrow \infty $ ),
the series in the second term of right hand side of (33) converges
to a finite sum, i.e.,
\begin{equation}
\displaystyle{ \delta_{N+1}^2  =
\left(\frac{\beta^2}{\alpha'^2}\right)^{N+1}\delta_0^2
+\frac{\alpha'}{\alpha'^2 - \beta^2}.}
\end{equation}

If $\delta_0$ is considered as arising out of Doppler broadening
only, we can estimate  $\delta_{\rm Dop}  \sim 10^9$  for $T = 10^4$
K. On the other hand, for anisotropic medium, using previous
works$^{10,40}$, \ we take \  $$\sigma_x = \sigma_y = 34.2{\rm \
nm}, \ \ \sigma_z = 8.731 \mu {\rm m} \ \ {\rm and} \ \ \sigma_\tau
= 1.3756
 \mu{\rm m}$$ and, \ the relevant components of the unit
 vectors \textbf{u} and ${\bf u}^\prime$ \ specified by \ $u_z = 0.8, \ u^\prime =
 0.9211 \ \ {\rm and} \ \  {\bf u}. {\bf u} = 0.965$, \  values of $\alpha$, \ $\beta$ \
 and \ $\alpha^\prime$ \ are:
$$\alpha' = 6.4608\times 10^{-28}s^2, \ \alpha = 5.6389 \times 10^{-28}s^2,
\ \beta = 7.4065 \times 10^{-28}s^2 \ \ {\rm for} \ \ \theta =
{15^0}$$ \noindent The Eq.(36) points to the fact that the second
term will be much larger than the first term, and effectively,
Doppler broadening becomes comparable with that due to multiple
scattering effect which, however will be heavily dependent on the
nature of the spatio-temporal fluctuation of the intrinsic physical
parameters within the medium. \vskip5pt \noindent

Next, for the opposite condition, i.e., $~~\al'~<~|\beta|$, the
series in (33) will be a divergent one and $\dl_{N+1}^2$ will be
finitely large for large but finite $N$. However, if the condition
(35) is to be satisfied, then the shift in frequency will be larger
than the width of the spectral lines. In that case, the condition
$\alpha'~<~|\beta|$ indicates that blueshift may also be observed.
On the other hand, the width of the spectral lines can be large
enough depending on how large the number of collisions is.
Therefore, in general, the blueshifted lines will be of larger
widths than the redshifted lines and may not be as easily observable
i.e., effectively, there will be blurring. \vskip10pt \noindent Now,
to find out the critical source frequency above which spectrum can
be analyzable, let us take the first equation of  equation(35),
which, after rearrangement gives
 \beq \left(\dl_n-\frac{\om_0}{2}\right)^2 \ll
\frac{\om_0^2}{4}-\frac{\al'}{\al\al'-\bt^2} \eeq \noindent Since
 the mean frequency of any source is always positive, {\it i.e.,}
$\om_0 \geq 0$, we must have \beq \om_0 \gg
\sqrt{\frac{4\al'}{\al\al'-\bt^2}}. \eeq By giving the right side of
this inequality the name  {\bf \it critical source frequency } ${\bf
\om_c}$, we obtain:
 \beq \om_c = \sqrt{\frac{4\al'}{\al\al'-\bt^2}}.
 \eeq
 Thus for a particular
medium, between the source and the observer, the  critical
 source frequency is the lower limit of the frequency of any
source whose spectrum can be clearly analyzed. In other words, the
shift of any spectral line from a source with frequency less than
the critical source frequency for that particular medium cannot be
detected due to its high broadening. \vskip5pt \noindent
 Now, we, arrive in a position to classify the light of the source depending
on the surrounding environment it passes through, on the way to the
point of observation, i.e., after passing through a scattering
medium characterized by the parameters $\al$, $\al'$, $\bt$. If we
allow only small angle scattering in order to get prominent spectra,
according to our DMS mechanism,  they will  either be blueshifted or
redshifted. The redshift of spectral lines may or may not be
detected according to whether the condition (38 \& 39) does or does
not hold. In this way, those sources whose spectra
are redshifted, are classified in two cases, i.e.,  \\
\vskip3pt \noindent
(i) \ \ \ $ \om_0~~>~~\om_c$ \ \ \ and  \ \ \ (ii) \ \ \  $\om_0~~\leq~~\om_c$. \\
\vskip3pt \noindent In the first case, the shifts of the spectral
lines can easily be detected due to condition (38). But in the later
case, the spectra will suffer from the  resultant blurring.

\section{\bf Role of Surrounding Environment in Quasar Astronomy:
Dynamic Multiple Scattering(DMS)}
 \noindent  It should be noted that the above framework of Dynamic
Multiple Scattering(DMS) for redshift measurement is applicable to
cosmological scales only when  the  environment around sources of
astronomical distances (say, quasars) is considered to have a random
medium with weak fluctuations due to the processes in and around the
source through which the scattered light travels to the observer.
For the time being, let us consider that the absorption cross
section is negligible compared to that of scattering. What we
conclude about the nature of cosmological sources is mainly through
spectra obtained by many means such as  X-ray, optical, infra-red
etc. observations i.e, using different types of astronomical tools
based on different wavelength ranges and selection criteria. The
light waves originating from the source, reach us after passing
through a vast amount of space with numerous physical interactions,
carrying the signatures not only of their own nature but also
allowing us to have a glimpse of the cosmological history of
evolution. \vskip5pt \noindent
\subsection{\bf Non-linearity of Hubble Law towards High redshift ($z\geq 0.3$)}
\subsubsection{\bf Statistical Point of View:}
\noindent Following similar techniques, as employed by Efron et
al.$^{41,42}$, in their recent works, Roy et.al.$^{43}$ analyzed the
relation between redshift and apparent magnitude from the
V\'eron-Cetty quasar catalogue(2006)$^{43}$(Fig.1a), using
statistical techniques well suited for truncated data.

\begin{figure}[h]
      \centerline{
    \epsfig{file=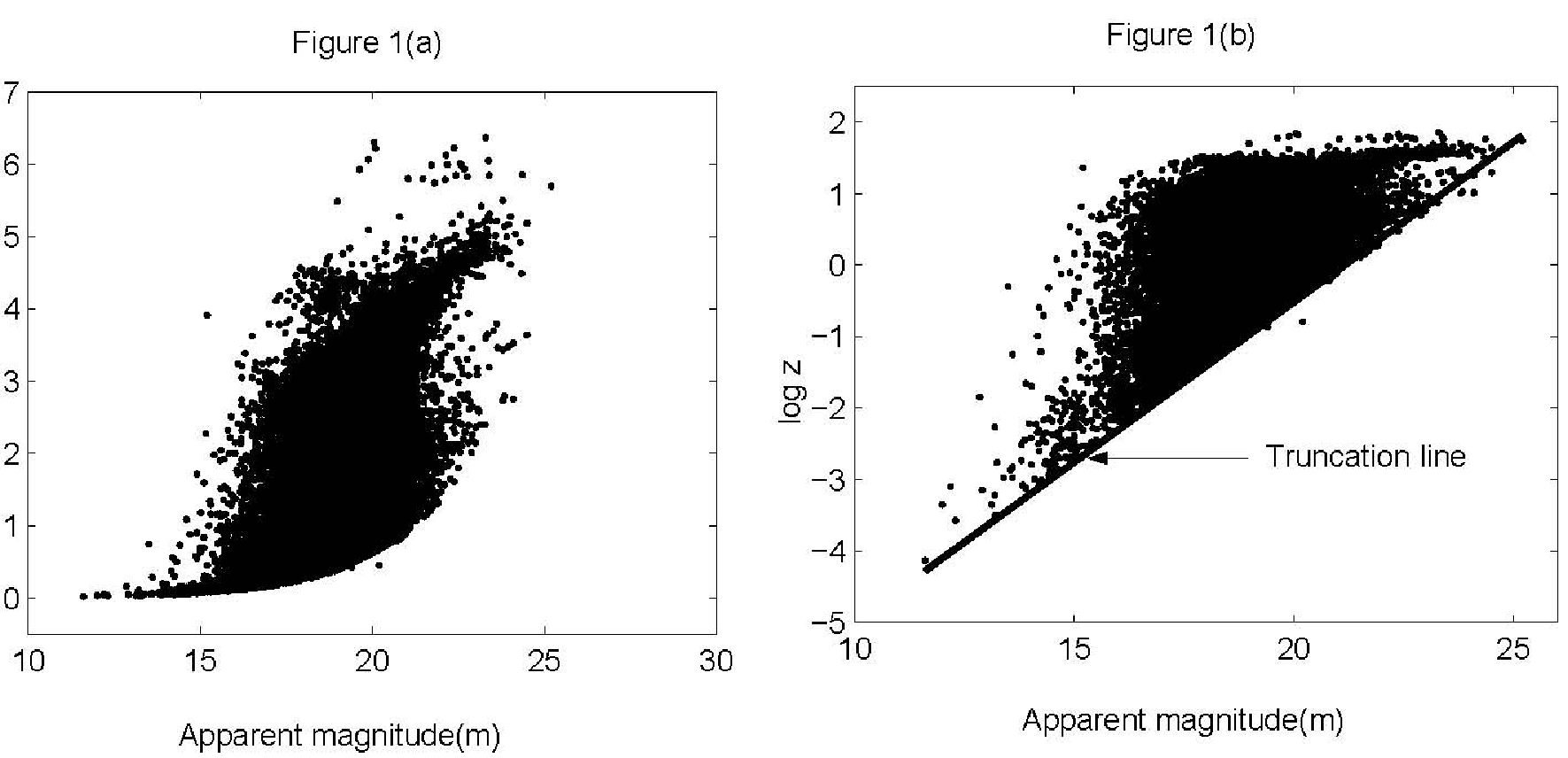, scale=.80}
    }
 \caption{\footnotesize{Apparent magnitude $(m)$  versus redshift$(z)$ in linear
scale$(a)$ and in log-scale$(b)$ with truncation line, from
V\'eron-Cetty(V-C) $(2006)$ quasar catalogue.}} \label{fig-eg1}
\end{figure}

The magnitude limited survey employed in that catalogue points to
the diagonal truncation boundary as is evident in fig.$1(b)$. The
analysis clearly shows
 the redshift data for the range \ $0 < z < 0.295$ \ fitted
quite satisfactorily with a linear diagram of \ $logz$ \ in \
apparent magnitude ($m$) vs. redshift \ ($z$) \ diagram, agreeing
well with the Hubble diagram. However, for the range \ $0.3 <z <
\simeq 3$, \ the data can be fitted with certain degree of nonlinear
dependence, showing spread of redshift corresponding to a single
value of apparent magnitude or vice versa. \vskip5pt \noindent
\subsubsection{\bf Doppler Shift vs. Shifting due to Dynamic Multiple
Scattering} \noindent  As discussed above, the current astronomical
observations point to the presence of the environment around quasars
of diverse nature. So, for higher redshifts, some other mechanism
should be effective and might play the dominant role compared to the
Doppler mechanism. We  propose the significant role of environment
in our understanding of the redshift of quasars, especially through
the effective evolutionary phase of their life.
This can occur in the following manner : \\
We know
$$(1+z) = (1+z_{\rm DMS})(1+z_{\rm D})$$
which gives,
$$z = (1+z_{\rm DMS})(1+z_{\rm D}) - 1$$
$z_{\rm DMS}$ being the redshift due to DMS mechanism and $z_D$ is
that due to Doppler mechanism.
\begin{figure}
       \centerline{
     \epsfig{file=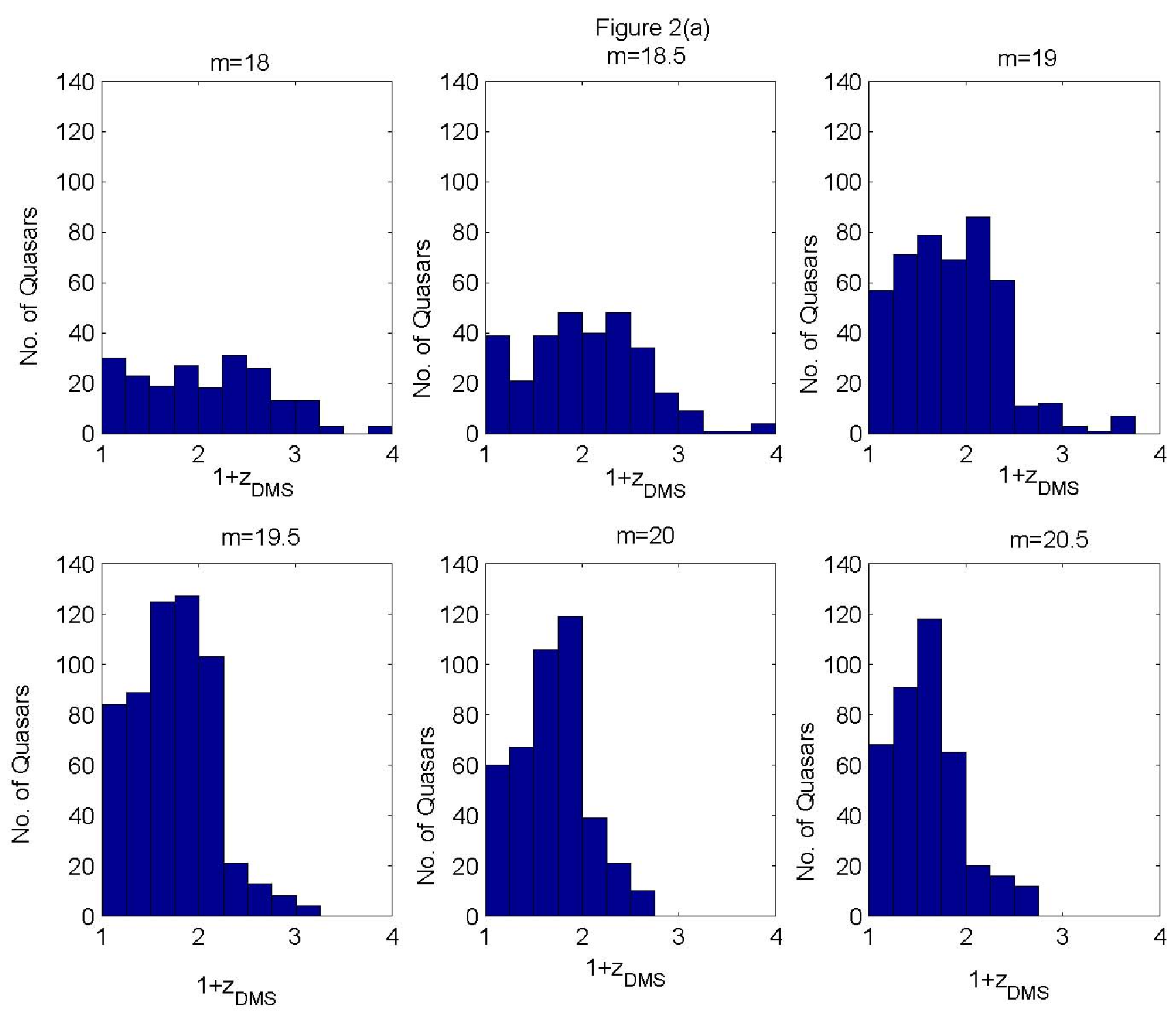, scale=.60}
     }
%% \caption{}
\label{fig-eg1}
\end{figure}
\begin{figure}
       \centerline{
     \epsfig{file=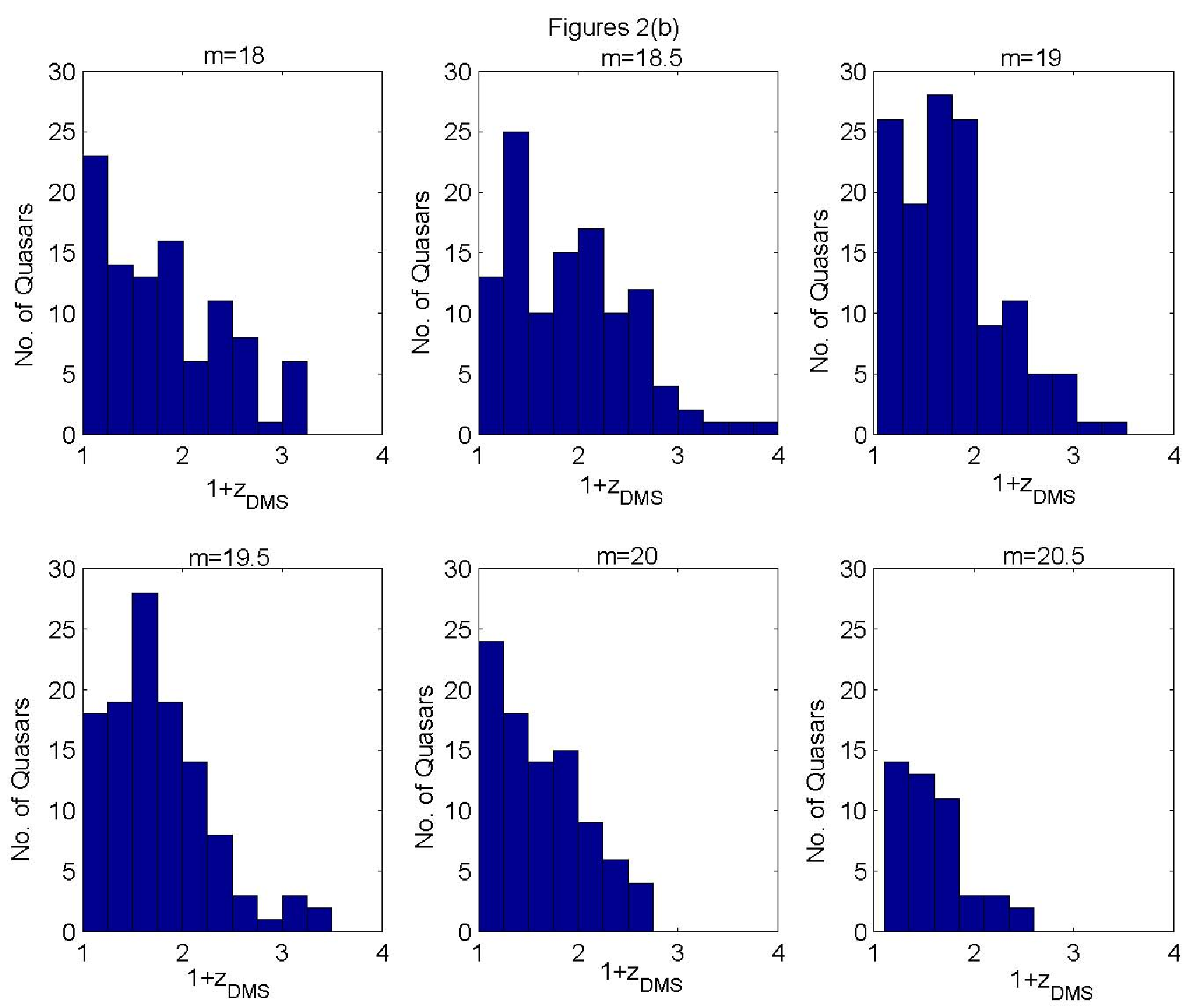, scale=.60}
     }
  \caption{\footnotesize{ Relation between number of quasars and the redshift from
V-C$(2006)$ quasar catalogue showing the contribution $z_{\rm DMS}$
from induced correlation mechanism(Wolf Effect) to redshift $(z)$
through the correlation parameters \ $\alpha, \alpha^{\prime}$ and
$\beta$ \ of the medium. Contribution is more prominent in the case
of Radio-quiet quasars$(a)$ than that in Radio loud quasars$(b)$.}}
\label{fig-eg1}
\end{figure}
Now, let us take a particular value of $m$ and the corresponding
redshift $z$ which lies on the truncation line as shown in
Fig.$1(b)$ and, also, assume the $z$ values lying in the truncated
line as that due to Doppler mechanism (as a rough approximation)
whereas the spread in $z$ for a particular $ m$ is mainly that due
to the environmental effects. Then, considering that DMS active in
these type of environments, we obtain from the equation(31),
\begin{equation}
1+ z_{\rm {DMS}} = \frac{\alpha^\prime +
 (\alpha \alpha^\prime -
\beta^2)\delta^2_ N}{|\beta|}
\end{equation}
where,
$$\delta^2_ N = (\frac{\beta^2}{{\alpha^\prime}^2})^ N \delta_0^2
+ \frac{\alpha^\prime}{{\alpha^\prime}^2 - \beta^2}$$ \noindent If
$\delta_0$ \ is considered as due to Doppler broadening only, then
$$\delta_ N^2 \simeq \frac{\alpha^\prime}{{\alpha^\prime}^2 -
\beta^2}$$
Therefore,
\begin{equation}
1+ z_{\rm {DMS}} = \frac{\alpha^\prime + (\alpha \alpha^\prime -
\beta^2)[\frac{\alpha^\prime}{{\alpha^\prime}^2 -
\beta^2}]}{|\beta|} \end{equation}
 Then,  what we observe as  redshift, is
\begin{equation}
z_{obs} = \frac{\alpha^\prime}{\beta} \left [ 1 + \frac{(\alpha
\alpha^\prime - \beta^2)}{{\alpha^\prime}^2 - \beta^2} \right ][1 +
z_{D}]-1
\end{equation}
or,
\begin{equation}
1 + z_{\rm obs} = \Gamma(1 + z_{D})
\end{equation}
where $\Gamma$ \ is a constant, dependent solely on the correlation
parameters $\alpha, \ \alpha^\prime, \ \beta$ of the medium through
which the wave comes to the observer after experiencing multiple
scattering.

The scatter plot of $m$ versus $z$ (Fig.$1(a)$)shows
that there exists a set of $z$ even for a fixed $m$ i.e., a set of
quasars, probably for diverse values of  $\alpha, \ \alpha^\prime, \
\beta$. \vskip5pt \noindent Taking the values of $z$ \ as \ $z_{\rm
D}$(${\rm D}$ for Doppler) \ from the truncation line(shown in
Fig.$1(b)$), we find $(1+z_{\rm obs})$ \ from the eq.(43). Assuming
a set of values for $\alpha, \ \alpha^\prime, \beta$, and
corresponding values of $\Gamma$ which can produce a certain extra
redshift other than $z_{\rm D}$, \ are obtained. Number of quasars
 vs. for a particular value of $z_{\rm D}$ \ has been plotted in Fig.$2(a \& b)$,
\ for various $m$, separately for radio quiet(Fig.2a) and radio loud
quasars(Fig.2b).

We get $\Gamma \simeq 2.06$ which is solely dependent on the nature
of the environment created in the laboratory condition. It is clear
from the bar-diagrams(Fig. $2(a \& b)$) that there exists a sizable
number of quasars around $\Gamma \simeq 2.06$ which implies that
there exist large number of quasars for a particular value of
$\Gamma$, indicating similar type of environments.
\vskip5pt\noindent
\subsection{\bf Effects on Distance Modulus and Line width}
\noindent At this point, we can conclude that the intrinsic physical
parameters of the medium (density, temperature , pressure and
ionization characteristics, the effect of which are manifested
through the induced correlation parameters as stated above) \
through which the source radiation crosses vast astronomical
distances to the point of observation, can contribute to the
shifting as well as to the width of a spectral line. We can then
write the relation between the width and the shift as[details in
reference$^{(8)}$]
\begin{equation}
W = \kappa {( \epsilon^2 + \eta^2 z^2 )}^\frac{1}{2(1+z)}
\end{equation}
\vskip5pt \noindent where $\kappa$, $\epsilon$ and $\eta $ are
constants. $\kappa\epsilon = \delta_0^2$, \ $\delta_0^2$ \ being the
 intrinsic spectral width of the source frequency. Taking the logarithm of both
 sides we can write
\begin{equation}
ln W = \frac{1}{2(1+z)} ln \kappa ( \epsilon^2 + \eta^2 {z}^2)
\end{equation}
It is well known $^{(40)}$ that the distance modulus $d$  can  be
written in terms of redshift $z$ as
\begin{equation}
 d = m - M =  42.38 - 5 ln (\frac{H}{100}) + 5 ln z + (1.086)(1-q_0)z
+ O(z^2)
\end{equation}
Where $H$ is the Hubble constant in km/sec/Mpc and $M$ refers to the
absolute magnitude. For small z i.e. $ z <<1$ , the eq.(46) reduces
to

$$d = 42.38 - 5 ln(\frac{H}{100}) + \frac{1.086}{2} z \ \ \ {\rm for} \ \ \ q_0 =
 \frac{1}{2}$$
 \noindent
This, after simplification becomes
$$z = \frac{d-C}{0.543}, \ \ \ {\rm where} \ \ \ C = 42.38 - 5 ln\left(\frac{H}{100}
 \right)$$
Substituting this value of $z$ \ in eq.(45), we have
\begin{equation}
d = - \frac{1.086}{ln E} lnW + N \ \ \ {\rm where}, \ \ \ ln E =
ln(\kappa \epsilon^2) \ \ {\rm and} \ \ \ N = C + 0.543
\end{equation}
 \vskip5pt \noindent The above relation(eq.47) between the
distance modulus and the width has a striking similarity to the
Tully-Fisher relation$^{(45)}$ but without any angular dependence.
The reason is obvious since we have considered the shift and width
due to scattering only, without considering any rotational effects
(in other way, the angle of scattering is considered as very small
i.e., $\theta \leq 15^0$ which is average in case of quasar). It
appears that in the case of photons which are emitted almost
perpendicular to the plane of the galaxy, we will be observing those
photons only without any rotational effects.
\subsection{\bf Effects on Hubble Flow at High Redshift($z \geq 0.3$)}
\noindent
 The extension of the redshift($z$) - apparent magnitude($m$) diagram towards high redshift
 in the Hubble diagram can be written as$^{46}$,
$$m = M - {\mathbf{K}}(z) - E(z) + 5ln cq_0^{-2} \left \{ q_0 z + (q_0-1)[(1+2q_0 z)^\frac{1}{2} - 1] \right \} + {\rm constant}$$
where $m$ is the apparent magnitude, $M$ is absolute magnitude.

$E(z)$ is a correction term due to evolution which calculates the
change in the luminosity of the main sequence termination point in
the Hertzsprung-Russel diagram for stars in a standard elliptical
galaxy, as a function of time. However, following Sandage$^{47}$ ,
the value adopted here, is
 $$E(z) \approx ln \left (\frac{1}{1+z} \right )$$
 ${\mathbf{K}(z)}$ is the  ${\mathbf{K}}$-correction term which accounts for the
 effect of $z$, i.e., it takes into account the fact that the galaxy is no
 longer being observed at a particular wavelength, but at what for small redshift($z$),
 it would be a shorter wavelength, and also with a bandwidth term $2.5 ln(1+z)$
to account for the stretching of the spectrum. Thus, ${\mathbf{K}}$
can be written as
 $${\mathbf{K}}= -2.5 ln(1+z)^{1-\alpha}, \ \ \ \alpha = 0.3$$
 \noindent
 assuming the spectrum of quasars varies as $\nu^{-\alpha}$.
Through $q_0$, termed the deceleration parameter, we must take into
account the effects of curvature. However, there is problem
considering simultaneously the effects of both the Evolution
parameter and different values of $q_0$ which measures the departure
of the linear Hubble law i.e., $m-ln cz$ linear relation, generally
in good arrangement for the low-redshift limit only. Following
specific cosmological model i.e., with $q_0=1/2$ for Einstein-de
Sitter model, we obtain,
\begin{equation}
m = M - {\mathbf{K}}(z) - E(z) + 5ln cz(1 + z/4) + {\rm constant} \
\ \ \ \ {\rm for} \ \ \ q_0 = 1/2
\end{equation}

\noindent Using eq.$(47)$ and $(48)$, the relation for apparent
magnitude and absolute magnitude for standard model, i.e., $q_0 =
1/2$ can be written as
\begin{equation}
m = A + B + E + {\mathbf{K}} - C*D
\end{equation}
where, \\
$A = m - M+{\rm constant}$ = $21.4$ \ is a constant term; \\
$B = \frac{5}{2.303}* lncz(1+\frac{z}{2})$ \ for $q_0 = \frac{1}{2}$ \\
$C = \frac{0.543}{(1+z)}$ \\
$D = 1+\frac{k}{2.303}* ln(1+F*z^2)$; \\
$E = $ \ Evolution term; \\
${\mathbf{K}} = - \frac{2.5}{2.303}(1+z)^{1-\alpha}; \ \ \ {\rm for} \ \ \ \alpha=0.3$ \\

\begin{figure}
       \centerline{
     \epsfig{file=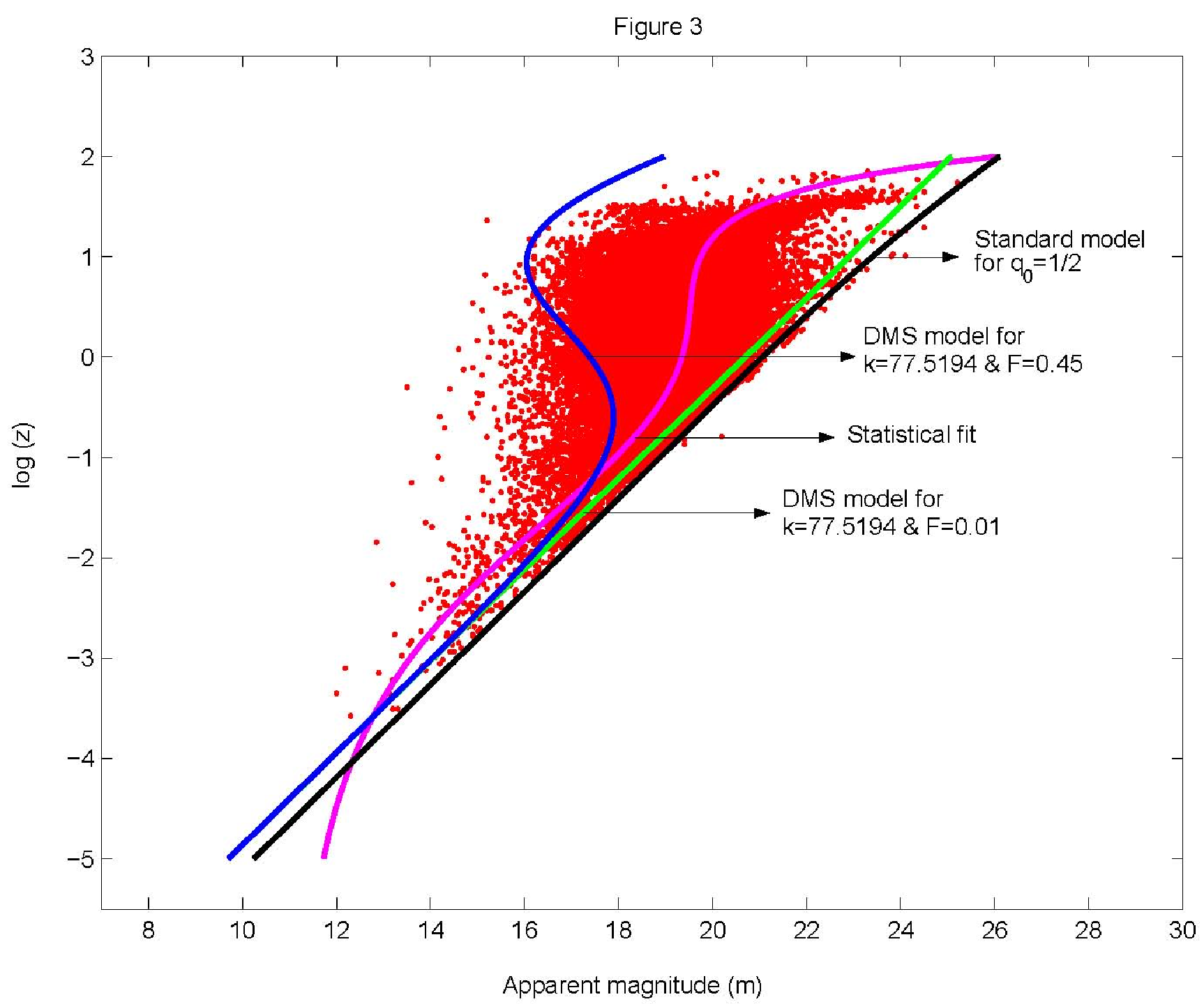, scale=.60}
     }
 \caption{\footnotesize{Comparison between Hubble curve with Standard
cosmological model($q_0=1/2$), considering V-C$(2006)$ quasar data
with that considering the proposed local environmental effects due
to Induced correlation mechanism present in the medium together with
statistical fitting.}}
 \label{fig-eg1}
\end{figure}
\noindent $C$ and $D$ being  the contributions due to the multiple
scattering effect, as calculated from of induced correlation
mechanism proposed in our framework, with $k = 1/ln(\epsilon)$ \ and
\ $F=\eta/\epsilon$. Taking the value of $F$ \ and \ $k$, we
calculate corresponding Hubble diagram with the standard model i.e.,
for \ $q_0 = 1/2$. But as we adopt a range of values for $F={0.45,
.... 0.01}$, \ we obtain a {\it bulge} \ in Hubble diagram when
approaching higher redshift limit(Fig.3). The limit of the values
$k$ \ and  $F$ \ as considered here, is intrinsically related to the
induced correlation parameters of the medium surrounding the
quasars. It is worth mentioning that, very recently, Cheng,
Permutter$^{22,23}$ reviewed in detail the current situations,
regarding Supernovae-redshift measurements and the corresponding
Hubble diagram. They explored the possibility of having certain but
variable values for cosmological parameters like $\Omega$ and the
cosmological constant($\Lambda$) in order to consider the idea of
accelerated expansion together with a phase transition in the
evolution history of the universe and proposed an explanation of the
{\it bulging} Hubble curve. \vskip5pt \noindent  Now, from eq.(45),
for $z = z_1$, \ the shift after the first scattering, is
$$z_1 = \frac{ln \epsilon}{2 \eta/\epsilon}$$
\noindent By taking, say, $F= 0.5$, and $k= 1/ln\epsilon \simeq 78$,
the contribution in our framework from first scattering is $\sim
0.02$. It is interesting to note that the experimental values,
mentioned by James and Wolf$^{10}$, for $\alpha^\prime/\beta$ \ is
found to range from 0.0085 to 0.146.
\section{\bf Possible Implications}
\noindent \vskip5pt The above analysis clearly indicates the
following possibilities:
\begin{enumerate}

\item  The DMS theory is applicable to a class of random media,
similar to those which may be present around quasars especially as
we proceed from low redshifts ($z \leq 0.295$), gradually, towards
higher ranges of $z$, and this extra shift can be explained without
even considering the relative motion between the observer and the
source or the medium.

\item  The evolutionary effect of this medium will be reflected in
the DMS through a set of parameters $(\alpha, \alpha^\prime ,
\beta)$ \ which are intrinsically related to the different physical
characteristics of the media such as general dielectric
susceptibility($\eta$). This, in turn is directly dependent on other
physical parameters such as the refractive index($n$),
permeability($\epsilon$), conductivity, temperature, pressure as
well as the molecular content and degree of ionization due to the
presence of the electromagnetic field.

\item  However, the amount of the contribution from DMS will be
dictated by the degree of the randomness or turbulence present in
the medium. Here, the degree of the fluctuation of the above
mentioned parameters should be such that the medium can be treated
in the weak fluctuation case. We expect our mechanism to be more
effective for radio quiet than for radio loud quasars (Fig. 2(a) \&
2(b))where we expect suitable environment or in other words, we can
describe the medium as weakly fluctuating. This information,
however, will be very important, when it will be more and more
available from astronomical observations(HST, KECK etc.). However,
quantitative studies should be developed for this purpose with
greater detail.

\item  The discrepancy in  the observed value of the quasars redshift should
    be traced back to both the Doppler as well as to the other possible
    local effects as reflected in DMS mechanism.
\end{enumerate}
\vskip5pt \noindent
 The clear deviation of the linear Hubble law for
high redshift ($z$) quasars is provided in our framework for a
certain range of medium parameters i.e., \ $\alpha$, \
$\alpha^\prime$ \ and \ $\beta$. This kind of deviation (or the
existence of a {\it bulge}) has already been mentioned by several
authors in connection with the Hubble diagram for high redshift
Supernovae$^{22,23}$. It clearly points to the fact that there is a
possibility of environmental effects on the observational aspects of
quasar-like astronomical objects  and their role should be
considered in explaining the observed bulge and associated other
criteria, i.e., the Hubble relation, especially at higher redshift
in addition to the Doppler effect or gravitational redshifts. In our
approach, we do not need to introduce any kind of phenomenological
approaches except to consider the role played by the characteristics
of the medium surrounding the radiating source.
 \vskip5pt \indent According to the currently established standard
models of quasars, spectral lines are emitted from the broad line
regions(BLR) as well as from the narrow line regions(NLR), i.e.,
from two types of clouds, named after the characteristics of lines
and with respect to the position of those clouds relative to the
torus surrounding the central black hole (central engine). Due to
the gravitational attraction, matter from the dust torus is heated,
causing it to radiate with a broad, non-thermal power spectrum. This
radiation excites gaseous clouds which emit line radiation. It is
worth mentioning that Wold et al.$^{48}$ have made a survey of
quasar environments at $0.5\leq z \leq 0.8$ and also concluded that
the quasars are located in a variety of environments, whereas,
Hutchings$^{(49)}$, after his recent observations pointed out that
the environment that triggers quasars activity almost certainly
changes with cosmic time and with the nature and the characteristics
of the emission line gas around at higher redshift quasars. Again,
Papadopoulos et al.$^{50}$ reported the discovery of large amounts
of low-excitation molecular gas around the infrared-luminous quasar,
APM 08279+5255 at z = 3.91 and confirm the presence of an extended
reservoir of molecular gas with low excitation, $10$ to $100$ times
more massive than the gas traced by higher-excitation observations.
This raises the possibility that significant amounts of
low-excitation molecular gas may lurk in the environments of
high-redshift ($z > 3$) galaxies. Botoff and Ferland$^{51}$ and his
coworkers also  point out the possible consequences about the role
of turbulence and magnetic field in determining the line profiles of
the emission lines which ultimately effects redshift values.
 \vskip5pt
\indent Serious controversy already exists about the influence of
individual quasars on their environments. It has been claimed as
incontrovertible evidence for the evolution of $Ly\alpha$, \ lines,
especially in case of luminous quasars. From this, it appears that
there is a certain kind of distribution in number density of lines
and hence equivalent widths which hints to the evolution of the
hydrogen column density of individual clouds of probable primordial
origin, spread throughout the space. All together, it points to the
essential role of the medium which is faced by the light waves that
cross through the vast astronomical distance by experiencing
multiple scattering in their journey. It is the nature of different
physical parameters of the medium which determines ultimately the
shape and nature of the spectrum we receive. This has great
implications on many other aspects of astronomical observations, in
scattering phenomena where the suitable physical characteristics are
available for producing similar effects which we expect to study in
future. \noindent
\subsection{\bf Concluding remarks}
Our analysis is not complete for several reasons: At this time, we
are not ready to claim  adequate knowledge about the properties of
the medium (such as $\alpha$, \ $\alpha^\prime$ \ and \ $\beta$) of
the quasar environment. Also, as  we are not in a position to
specify quantitatively the underlying physical nature of the
scattering medium, other than its anisotropic coherence properties,
our model can only be considered as indicative of some
possibilities. The scatterer, which is assumed to have a ``white
noise'' power spectrum (implying that its fluctuations are very
energetic and anisotropic), is situated farther away from the center
of the radiating source than the line emitting clouds. These clouds
can not be too hot or otherwise they would be completely ionized,
making line radiation impossible. Besides these, so far we have
confined our discussion to situations where the scattering medium is
considered as weakly fluctuating medium and, also, if ionized,
sufficiently under dense so that it can be treated as underdense
plasma. Also, we have ignored the unscattered and backscattered
radiation in the present problem. This effect is manifested by
enhanced back scattering of light from disordered media and arises
from constructive interference between propagation along forward and
reverse paths. Lagendijk$^{52,53}$ noted that there is a connection
between coherence-induced spectral changes and the phenomenon of
weak localization. In this case, the spectral degree of coherence
does not satisfy the scaling law, and, consequently, the spectrum of
the backscattered radiation may be shifted relative to the spectrum
of the field, very much like Wolf effect.  \vskip20pt \noindent {\bf
Acknowledgements}
 \vskip15pt \noindent The authors
(S.Roy and M.Roy) are greatly indebted for the kind hospitality and
financial support from College of Science and Center for Earth
Observation and Space Research, George Mason University, USA, during
the period of this work. \vskip15pt \noindent {\bf References}
\vskip5pt \noindent
\begin{enumerate}
\item  Datta S., Roy S., Roy M. and Moles M.(1998), Int.Jour.Theort.Phys.,{\bf 37},N5, 1469.
\item  Datta S., Roy S.,Roy M. and Moles M.(1998), Phys.Rev.A,{\bf 58},720.
\item  Wolf E.(1986), Phys.Rev.Lett.,{\bf 56},1370.
\item  Wolf E.(1987),Nature,{\bf 326},363.
\item  Morris G.M. and Faklis D.(1987),Opt.Commun.,{\bf 62},5
\item  Faklis D. and Morris G.M.(1988),Opt.Lett.,{\bf 13},4.
\item  Wolf E.(1991),NPL Technical Bulletin:October 1991,p1
\item  Roy S., Kafatos M. and Datta S.(1999), Phys.Rev.A,{\bf 60}, 273.
\item  James D.V.F.,(1989), Phys.Lett.A.,{\bf 140},213
\item  James D.F.V. and Wolf E.(1990),Phy.Lett.A,{\bf 146},167
\item  James D. and Savedoff M.P.(1990), ApJ.,{\bf 359},67.
\item  Wolf E. and James D.F.V.(1996),Rep.Prog.Phys.,{\bf 59},771
\item  James D.F.V.(1998),Pure Appl.Opt.,{\bf 7},959-970
\item  Shirai T. and Asakura T.(1995),J.Opt.Soc.Am.A,{\bf 12},N6,1354.
\item  Shirai T. and Asakura T.(1996),Optical Rev.,{\bf 3},N5,323
\item  Wolf E. and Foley J.T.(1989),Phys.Rev.A,{\bf 40},N2,579
\item  Tatarskii V.(1998),Pure Appl.Opt.,{\bf 7},953-957
\item  Datta S, Roy S, Roy M and Moles M(1998),Int. Jour.Theor.Phys.,{\bf 37},N4, 1313.
\item  Roy S., Kafatos M. and Datta S., Astron.Astrophys.,{\bf
       353},1134-1138
\item  Savedoff M.(1989), News l.Astronom.Soc.NY,{\bf 3},22-23
\item  Sulentic J.W.(1989), Astrophys.J.,{\bf 343},54-65
\item  Tai-Pei Cheng, http://www.umsl.edu/~tpcheng/AccUniv2/sld047.htm
\item  Perlmutter S. and Schmidt Brian P., astro-ph/0303428 and references there in.
\item  Wolf E.(1989), Phys Rev.Lett.,{\bf 63}, 2220.
\item  Nussenzveig H.M.((1972), {\it Causality and Dispersion relations}(New York: Academic)
\item  Watson  Kenneth M.(1969), J.Math.Phys.,{\bf 10},N4,688
\item  Watson Kenneth M.(1970), Phys.Fluids,{\bf 13},N10,2514
\item  Van Hove L\'eon(1954), Phys.Rev.,{\bf 95},N1,249;(1958),Physica,{\bf 24},404
\item  Glaubar R.J.(1962) in {\it Lectures on Theoretical Physics IV}(Univ. of Colorado
       Summer Institute for Theoretical Physics), eds. W.E.Britten,B.W.Downs
       and J.Downs,(Interscience,New York),p.571
\item  Davenport W.B. and Root W.L.(1958),{\it An Introduction to the Theory of Random Signals and
       Noise}(New York:Mc-Graw-Hill)(Reprinted by IEEE Press, New York, 1987)
\item  Wolf E., Foley J.T. and Gori F.(1989), Opt.Soc.Am.A,{\bf 6},1142.
\item  Van Cittert P.H.(1934), Physica,{\bf 1},201
\item  Zernike F.(1938), Physica{\bf 5},785
\item  Villars F. and Weisskopf V.F.(1954), Phys.Rev.,{\bf 94},N2,232-240.
\item  Allis W. et al.,(1963),{\it Waves in anisotropic Plasmas}(MIT Press, Cambridge),p13.
\item  Born M. and Wolf E.(1998), ``Principle of Optics'', 6th edition, Pergamon,
       Oxford.
\item  Ishimaru Akira(1997) in {\it Wave propagation and Scattering in Random Media}
       (IEEE/OUP Series on Electromagnetic Wave Theory)
\item  Ishimaru A.(1978),{\it Wave propagation and scattering in random media, Vol
       II}(Academic Press, New York),section 16.5.
\item  Foley J.T. and Wolf E.(1989),Phys.Rev.A.,{\bf 40},588
\item  Gori F.,Guattari G.,Palma C.,and  Padovani G.(1988) Opt.Commun,{\bf 67},1
\item  Efron B. and Petrosian V.,(1992), Astrophys J.,{\bf 399}:345-352.
\item  Efron B. and Petrosian V.,(1999),Journal of the American
       Statistical Association,{\bf 94},N447,824-834.
\item  Roy S. et.al.(2006), Astro-phys/0605356, accepted for publication
       in Procd. of {\it 6th International Workshop on "Data analysis in
       Astronomy"}, Erice, italy.
\item  VERONCAT-V\'eron Quasars and AGNs(V2006),{\it HEASARC Archive,
       ftp://cdsarc.u-strasbg.fr/pub/cats/VII/248/}.
\item  Tully R.B. and Fisher, J.R.(1977), A \& A, {\bf 54}, 661.
\item  Narlikar, J. V., (1983), {\it Introduction to Cosmology},(Jones \& Bertlett Publishers,Inc.,Boston).
\item  Sandage A. and Tamman G.A., (1983), in {\it Large Scale structure of the Universe, Cosmology,
       and Fundamental Physics}, (eds. S. setti \& L. Van Hove, geneva, ESO/CERN),127
\item  Wold M.,Lacy M., Lilje P.B. and Serjeant S.(2001), in {\it "QSO Hosts
       and their Environments", QSO environments at Intermediate
       Redshifts and Companions at Higher Redshifts}, eds. by M\'arquez eta al.,
       Kluwer Academic/Plenum Publishers.
\item  Hutchings John B.(2001), in{\it "QSO Hosts and their Environments", QSO
       environments at Intermediate
       Redshifts and Companions at Higher Redshifts}, eds. by M\'arquez eta al.,
       Kluwer Academic/Plenum Publishers.
\item  Papadopoulos P. et al.(2001), Nature, 409, 58.
\item  Botoff M. and Ferland G., Astrophys.J.(2002), {\bf 568},581-591.
\item  Lagendijk A.(1990a),Phys.Lett.A,{\bf 147},389.
\item  Lagendijk A.(1990b),Phys.Rev.Lett.,{\bf 65},2082.
\end{enumerate}

\newpage
 \vskip25pt \noindent
Figure 1:  Apparent magnitude $(m)$  versus redshift$(z)$ in linear
scale$(a)$ and in log-scale$(b)$ with truncation line, from
V\'eron-Cetty(V-C) $(2006)$ quasar catalogue.
 \vskip25pt \noindent
Figure 2 : Relation between number of quasars and the redshift from
V-C$(2006)$ quasar catalogue showing the contribution $z_{\rm DMS}$
from induced correlation mechanism(Wolf Effect) to redshift $(z)$
through the correlation parameters \ $\alpha, \alpha^{\prime}$ and
$\beta$ \ of the medium. Contribution is more prominent in the case
of Radio-quiet quasars$(a)$ than that in Radio loud quasars$(b)$.
\vskip25pt \noindent
Figure 3: Comparison between Hubble curve with
Standard cosmological model($q_0=1/2$), considering V-C$(2006)$
quasar data with that considering the proposed local environmental
effects due to Induced correlation mechanism present in the medium
together with statistical fitting.
\end{document}